\documentclass[12pt,preprint]{aastex}

\slugcomment{To appear in PASP (September 2003, accepted)}

\shorttitle{The Pisgah Automated Survey}
\shortauthors{L$\acute{o}$pez-Morales \& Clemens}

\begin{document}


\title{The Pisgah Automated Survey: A Photometric Search for Low-mass Detached Eclipsing Binaries and Other Variable Stars}


\author{M. L$\acute{o}$pez-Morales and J. Christopher Clemens\altaffilmark{1}}
\affil{Dept. of Physics and Astronomy, University of North Carolina\\
    Chapel Hill, NC 27599}
\email{lopezmor,clemens@physics.unc.edu}

\altaffiltext{1}{Alfred P. Sloan Research Fellow}


\begin{abstract}
The Pisgah Survey, located at the facilities of the Pisgah Astronomical Research Institute in Rosman NC, is a low cost project to acquire fully-automated I band photometry of selected areas of the sky. The survey collects multiple images of $\sim$ 16.5 sq. deg. of sky per night, searching for variability in stars with apparent magnitudes brighter than I $\sim$ 15. The main scientific goal of this project is to discover new low-mass detached eclipsing binaries to provide precise constraints to the mass-radius relation for the lower main sequence.

In this paper we present a technical description of the project, including the software routines to automate the collection and analysis of the data, and a description of our variable identification strategy. We prove the feasibility of our technique by showing the successful detection of the previously known M-dwarf detached eclipsing binary GJ 2069A, and we present the results of the analysis of the first set of fields imaged by the survey, in which 15 new variables have been discovered among 8,201 stars monitored. The paper concludes with an outline of the project's prospects.
\end{abstract}




\section{Introduction}

\begin{figure}
\plotone{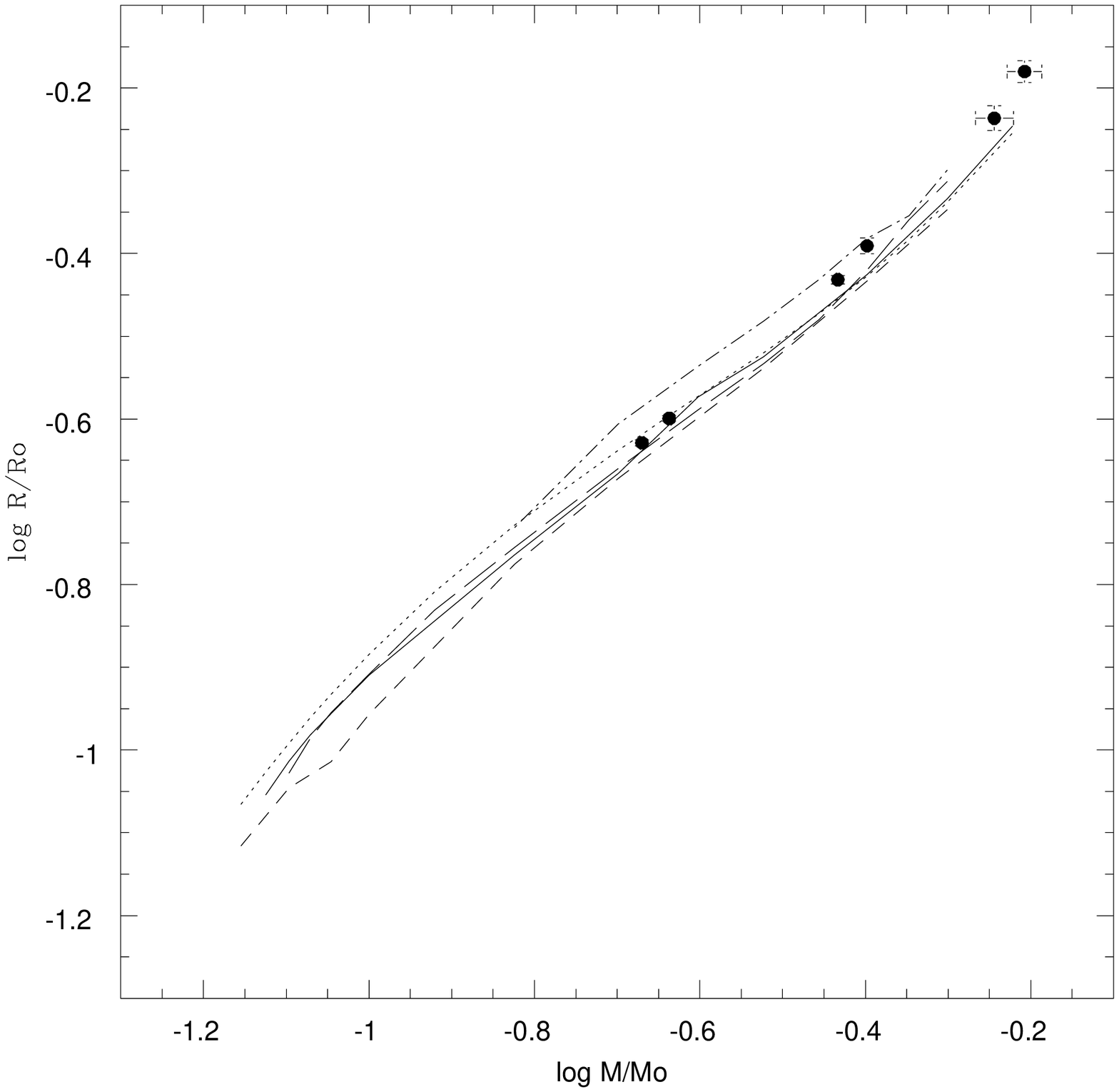}
\caption{Mass-radius relation in logarithmic scale showing all the current direct measurements of mass and radius for stars with masses below 1$M_{\sun}$. Those measurements, indicated by the dots in the figure, come from the detached eclipsing binaries YY Gem, GJ 2069A and CM Dra. The error bars for the last two binaries are smaller than the dots. The lines correspond to stellar models by \citet{bar98}, \citet{tou96}, \citet{dor89}, \citet{nee84}, and \citet{dan82}. \label{fig1}}
\end{figure}

The most fundamental observational parameters characterizing a star are its mass and its radius. Quantities like temperature, composition, and total luminosity cannot be measured without invoking model calculations, while mass and radius can be measured directly. Together they can provide rigorous constraints on models of stellar interiors. Ironically, we have more observational data on the mass-radius relation for planets (from our solar system) than we do for stars less than 1$M_{\sun}$. Figure~\ref{fig1} shows all direct measurements of mass and radius for stars with mass less than the Sun. These measurements come from the detached eclipsing binaries YY Gem \citep{bop74,leu78}, CM Dra \citep{lac77,met96}, and the recently discovered GJ 2069A \citep{del99,rib02}. This handful of stars is not sufficient to provide precision tests of theoretical mass-radius relationships.

In addition to the stars in figure~\ref{fig1}, there are many direct mass measurements of low mass stars from visual binaries (see Henry $\&$ McCarthy 1993) and recently \citet{seg02}, and \citet{lan01} have acquired direct radius measurements from interferometry. However, none of these measurements yields mass and radius simultaneously, nor are the radius measurements as precise as those measured from eclipsing binaries. The best way to improve the empirical mass-radius relationship is to discover new low-mass eclipsing binaries. This is most easily done by detecting their photometric variability, but the low number of known systems reflects the low probability of chance detection. Fortunately, the consumer market for digital imaging technology and for astronomical hardware and software has now made concerted searches for stellar eclipses both practical and inexpensive.

Improving the mass-radius relationship of low-mass stars, especially the M-dwarfs, is important and timely. Theorists have invested much effort in recent years to improve M-dwarf models. In particular, the state-of-the-art models by \citet{bar98} incorporate sophisticated Next-Gen atmosphere models \citep{hau99}, and now reproduce very well the mass-magnitude and color-magnitude relations of M-dwarf stars. However, these models still have some shortcomings which \citet{bar98} attribute to remaining uncertainties in the complex atmospheric physics of these cool objects; for the most massive M-dwarfs, for example, the models predict at solar metallicity bluer V-I colors than those observed.

These uncertainties in the atmospheric models prevent us from testing interior physics using color-magnitude and mass-magnitude relations. As explained by \citet{cha97}, the radius of their model stars is mainly fixed by the equation of state used in the interior, and it is only slightly affected by the atmosphere models. If their equation of state generates models that are too large or too small, the effect on predicted magnitudes is smaller than the effects of temperature ($R^{2}$ versus $T^{4}$), and could hide in the remaining uncertainties of the model atmospheres. Consequently, only an empirical determination of the mass-radius relation of the stars can yield stringent tests of the interior equation of state, which in the cores of late M-dwarfs includes untested corrections for Coulomb interactions.

The survey we describe in this paper is intended to discover new detached low-mass eclipsing binaries which can be used to add precision measurements to the mass-radius relationship. The survey is called the Pisgah Survey for Low-Mass Eclipsing Binaries, and it employs an 8-in aperture telescope and a 2048$\times$2048 CCD to perform fully-automated differential photometry of selected fields covering an area of $\sim$ 16.5 sq. deg. of sky every clear night. It monitors about 90,000 stars per year, contained in an area of 197.7 sq. deg. Besides finding new low-mass detached eclipsing binaries, the survey also detects and measures light curves of other periodic variables, especially those with periods of less than $\sim$ 1 month. The equipment can be accessed remotely, and the automated observing routine can be interrupted at any time for targets of opportunity such as gamma-ray bursts.

This paper describes the survey in detail, beginning in section 2 with an explanation of the criteria used to select the equipment and observing strategy. Section 3 presents the hardware and the software used to set up the system, and in section 4 we provide an overview of the routines developed to collect, reduce, and analyze the data to identify variable stars. Section 5  shows the results of the analysis of the first 14.76 sq. deg. of sky surveyed. Finally, section 6 summarizes the current status of the project and describes our future plans.

\section{Selection of the Survey Parameters}

We chose our survey parameters to optimize the detection of eclipsing binaries through their photometric variability. This required that we first understand the detection probability for various search strategies, and then that we balance this against the practical amount of observing time available and our modest budget for purchasing or constructing hardware and software. We also considered limits imposed by the photometric quality of the site available, and reasonable construction time scales. In this section we describe the rationale for our choices of hardware, software, and observing strategy.

\subsection{Monte Carlo Simulations}

The probability of detecting low-mass detached eclipsing binaries depends on intrinsic characteristics of those systems, such as their space density, their number distribution with mass (initial mass function), their apparent magnitudes, the ratios of sizes and temperatures of their components, their spatial inclinations, and the distribution of their orbital periods. The combination of the last four parameters determines the width and depth of the eclipses our survey seeks to detect. To be successful our equipment must conduct adequate time sampling of stars in a sufficiently large volume of space to make detection of new binaries a certainty. This requires an intelligent tradeoff between areal coverage of the sky, limiting magnitude for eclipse detection, and sampling rates.

To investigate these trades, we have developed a set of Monte Carlo simulations that incorporate our best knowledge of low-mass binary properties to estimate the number of low mass eclipsing binaries detected by a given instrumental setup\footnote{A more detailed description of the simulations and their results are being published in a separate paper \citep{lmcIP}}.

The simulations generate a sample of stars with the space density and mass distribution derived by \citet{rei97} using stars within 8pc. The program then assigns binary companions to 35$\%$ of those stars, where the companion mass is drawn according to the binary mass-ratio distribution observed by \citet{reiB97} (no triple systems or higher are included in our sample). In the absence of a well-measured period distribution for binaries below 1$M_{\sun}$, we assign orbital periods to the binaries using the distribution found by \citet{duq91} for G stars. The simulation then distributes the binaries in space with random inclinations, and with a space density that drops above and below the galactic disk following a two component exponential to account for the thin and thick disk components (see Gilmore, King, and van der Kruit 1990). The binary magnitudes are assigned by look-up from the mass-magnitude relations of \citet{hen93}, and the radii are calculated from the approximate mass-radius relation of \citet{nee84}, which we chose to minimize computation time.

Once the sample is generated, the program selects those binaries that will show eclipses based on their orbital inclinations, orbital periods, and the radii of the stars. The simulation discards interacting binaries, defined as those in which at least one component fills its Roche lobe. It also eliminates all the binaries with eclipses narrower than 5$\%$ of the orbital period, because detecting these would require over 300 observations, a number we decided a priori to be impractically large\footnote{Our experiments with period search routines show that at least 10 measurements during eclipse are required to detect the eclipsing binaries.  With random sampling, the probability of acquiring these 10 measurements exceeds 95$\%$ after 300 observations for eclipses wider than 5$\%$ of the orbital period.}.  The sample of synthetic binaries that remains represents an estimate of all the binaries that are detectable in principle. Specific choices of hardware will reduce this number, so the final step is to impose detection limits dictated by various selections of hardware and observing strategy. Note that we have not placed any upper limit on the orbital periods; instead these limits occur naturally, since the eclipse probability falls so rapidly with increasing period that practically no binaries with periods longer than about 10 days are expected.

\begin{figure}
\plotone{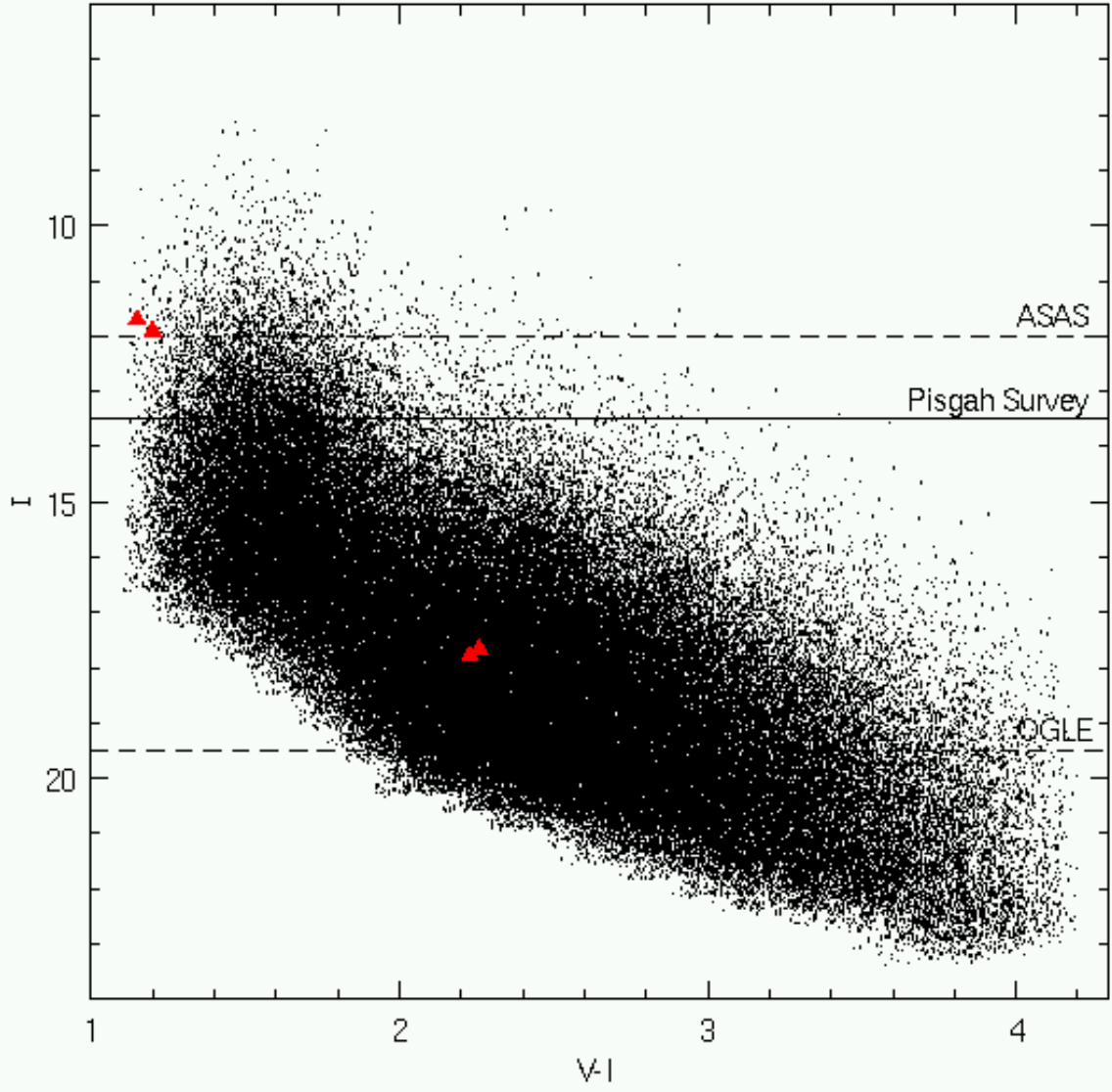}
\caption{I vs V-I color-magnitude diagram containing 131,348 simulated low-mass detached eclipsing binaries in the whole sky, as generated by the Monte Carlo simulations described in section 2. A detailed explanation of the features in this diagram is given in the text. \label{fig2}}
\end{figure}

\begin{figure}
\plotone{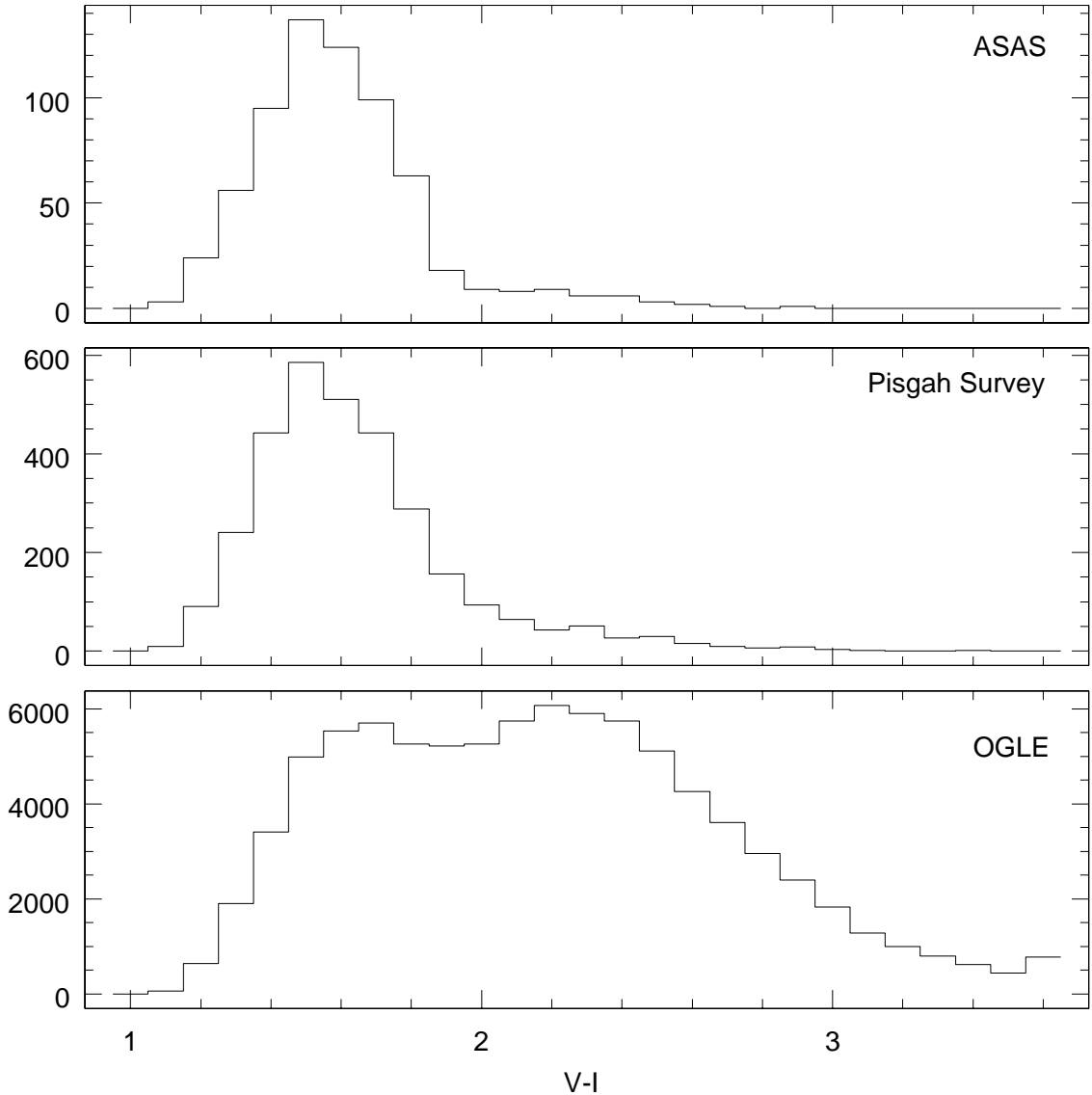}
\caption{Histograms representing the number distributions of binaries predicted by our simulations for the All Sky Automated Survey \citep{poj97}, the Pisgah Survey, and the Ogle Gravitational Lensing Experiment \citep{uda98}, respectively. From these histograms we conclude that a survey like ASAS will not be successful detecting M-dwarf eclipsing binaries, which have colors V-I $\geq$ 2.0. OGLE, on the other hand, should have no difficulty detecting M-dwarf systems, its only constraint been its small field-of-view. With a setup like the Pisgah Survey (see section 3 for a technical description) our simulations predict a discovery rate of 1 low-mass candidate brighter than I = 13.5 per each 30 sq. deg. of sky surveyed, after adjusting the detection rate to the results from OGLE. \label{fig3}}
\end{figure}

The results of the simulations are presented in figures~\ref{fig2} and~\ref{fig3}. Figure~\ref{fig2} shows an I vs. V-I color-magnitude diagram containing the 131,348 detectable low-mass detached eclipsing binaries predicted by the simulation in a search area covering the entire sky (this diagram is instrument independent). The vertical cutoffs in this diagram at V-I $\sim$ 4.2 and 1.37 result from restricting the masses in the simulations to be between 0.82 and 0.1 $M_{\sun}$. The cutoff at the top of the diagram reflects the declining volume of space with decreasing distances to the observer. The number of binaries increases with increasing magnitude in a manner that reflects the competition between increasing search volume, and the exponential decline in stellar space density associated with the two components of the Galactic disk. The number of detectable binaries decreases toward redder colors, in spite of an increase in the mass function, because the eclipses grow increasingly narrow as the stellar radii decrease. It is likely that this trend underestimates the real number of low-mass systems, because the binary period distribution we used in our calculations decreases monotonically at short period, while observations of nearby low-mass binaries \citep{rei97} and those in the Hyades \citep{duq91} show suggestions of an increase in numbers at short period.  Even a modest increase in the number of short period binaries can significantly change the number of detectable binaries at the low-mass end.  The two bands with slightly higher concentrations of binaries located between 1.4 $\lesssim$ V-I $\lesssim$ 1.9, and 2.1 $\lesssim$ V-I $\lesssim$ 2.7 reflect changes in the slopes of the magnitude transformation equations from \citet{rei97}, and the mass-Mv relation in \citet{hen93}. Both slope changes cause a large range of masses to accumulate in a narrow range of colors.

Superimposed in this diagram we show two horizontal dotted lines at I=12.0 and I=19.5, which represent the effective detection limits of two published photometric surveys we have used to calibrate our simulation\footnote{We set these limits equal to the faintest periodic variables detected in each survey}. Those surveys, both conducted by the Warsaw University Observatory in Poland, are the Ogle Gravitational Lensing Experiment \citep{uda92}, and the All Sky Automated Survey \citep{poj97}. Figure~\ref{fig3} shows the number distributions of binaries predicted by our simulations for those two surveys, after applying the corresponding survey dependent completion functions. These functions are necessary to model the dependence of detection on magnitude, since shallow eclipses can be detected in bright stars, but not in faint ones\footnote{The survey dependent parameters used to simulate those completion functions are the exposure time of the images, the magnitude detection limit of the survey, photometric completeness curves, and $\sigma_{I}$ vs. I curves (average magnitude dispersion of the stellar light curves vs. their apparent magnitudes)}. The decrease in numbers below V-I $\sim$ 1.4 is artificially steep in all the plots due to the upper mass limit we imposed.

In actuality, the Ogle experiment detected two foreground eclipsing M-dwarfs in a 2.4 sq. deg. field centered at the Small Magellanic Cloud \citep{uda98}, while ASAS found two new binaries with estimated masses of about 0.8 $M_{\sun}$ in its first phase covering 300 square degrees \citep{poj00}. The triangles in figure 2 show the estimated locations of these four systems. The ASAS detections are at the extreme upper end of the mass range we considered, so we cannot reliably compare them to the expected rate.  For the OGLE survey, the real detection rate is smaller by a factor of 2.5 than the expectation from our simulations, which may be only a statistical variation or simply indicate that the OGLE survey incompleteness does not follow the functions we applied to the data. Both OGLE detections were 1.0 to 1.5 magnitudes brighter than the magnitude limit, favoring this possibility.  Alternatively, the value we have used for the space density of low mass stars, 0.06 stars $\cdot$ $pc^{-3}$ \citep{rei97}, may be too high.  The number is based on known stars within 8 parsecs, and could be low if the sample is incomplete, but it is difficult to argue that it is too high.  It is even more unlikely that our simulation overestimates the number of short period binaries, since the 8 parsec sample actually contains 3 such eclipsing binaries while our simulation predicts none\footnote{Behind these numbers undoubtedly lies an interesting tale, since the three low-mass eclipsing binaries within 8 parsecs could represent a population of 20-30 such systems that would push the binary fraction in the 8 parsec sample much higher, and create a large spike at short period in the measured orbital period distribution.}. Whatever the explanation for the shortage of detections, our search strategy will be similar to that of the OGLE survey, so we should have a similar ratio between expectation and detection, and can scale the output of our simulation accordingly.

Figures~\ref{fig2} and~\ref{fig3} reveal that a survey with the  characteristics of ASAS will not likely detect M-dwarf eclipsing binaries, since the number of detectable systems in the whole sky drops to almost zero at V-I = 2.0. OGLE, on the other hand, has no difficulty detecting M-dwarf systems, but has only small areal coverage. Consequently, OGLE will mostly detect binaries too faint for practical spectroscopic follow-up.

We would like our survey limit to be bright enough that all the binaries detected will be candidates for spectroscopic follow-up on 4-meter class telescopes or smaller. At the same time we need to reach deep enough in magnitude to successfully detect M-dwarf binaries in a reasonable amount of sky. Our simulations show that a binary detection limit of I$\sim$13.5 is a reasonable value to meet those two requirements; below this magnitude the SNR on readily-available high resolution spectrographs becomes too poor to measure the binary's radial velocity curves accurately. Based on the ASAS survey, where the faintest variable is 1.0 - 1.5 magnitudes brighter than the survey's photometric detection limit, we estimate that detecting binaries with magnitude 13.5 will require a photometric detection limit of I$\sim$ 14.5-15.0 

The continuous horizontal line in figure~\ref{fig2} shows the results of the simulation for a binary detection limit of I = 13.5. With this limit we can detect systems down to V-I $\sim$ 3.0, which corresponds to primary masses of about 0.2$M_{\sun}$. The histogram of color distribution for all the binaries above this limit is shown in the central plot of figure~\ref{fig3}. There are about 3122 detectable binaries in the whole sky and 202 binaries with V-I $\geq$ 2.0 in this histogram.  Thus we expect that a survey with a detection limit of I = 13.5 will find one low mass detached eclipsing binary in every 13 $deg.$ surveyed and one M-dwarf binary in every 204 sq. deg. of sky surveyed. If we adjust the detection rate downward, based on OGLE detections, these numbers become 32.5 and 510 degrees. This is an acceptable yield for a dedicated telescope, and we shall see in the next section that the requirement of 15th magnitude photometry is easy to satisfy using equipment from the high-end consumer market for astronomical hardware.

\section{Hardware}

In this section we describe the observing tools we have built to accomplish our survey. In addition to repeatedly imaging large areas of the sky to a magnitude limit of I = 15, we wanted the survey to require modest construction time and cost, and, once working, minimum human intervention during operations. With these goals in mind we assembled our instrument using, whenever possible, off-the-shelf commercial parts for both the hardware and software components. Not all of our choices turned out well, but after several phases of troubleshooting and trial-and-error tests, we arrived at the fully operational system summarized in Table~\ref{tbl-1}, and described in more detail below.

\subsection{Telescope and CCD system}

\begin{figure}
\epsscale{0.8}
\plotone{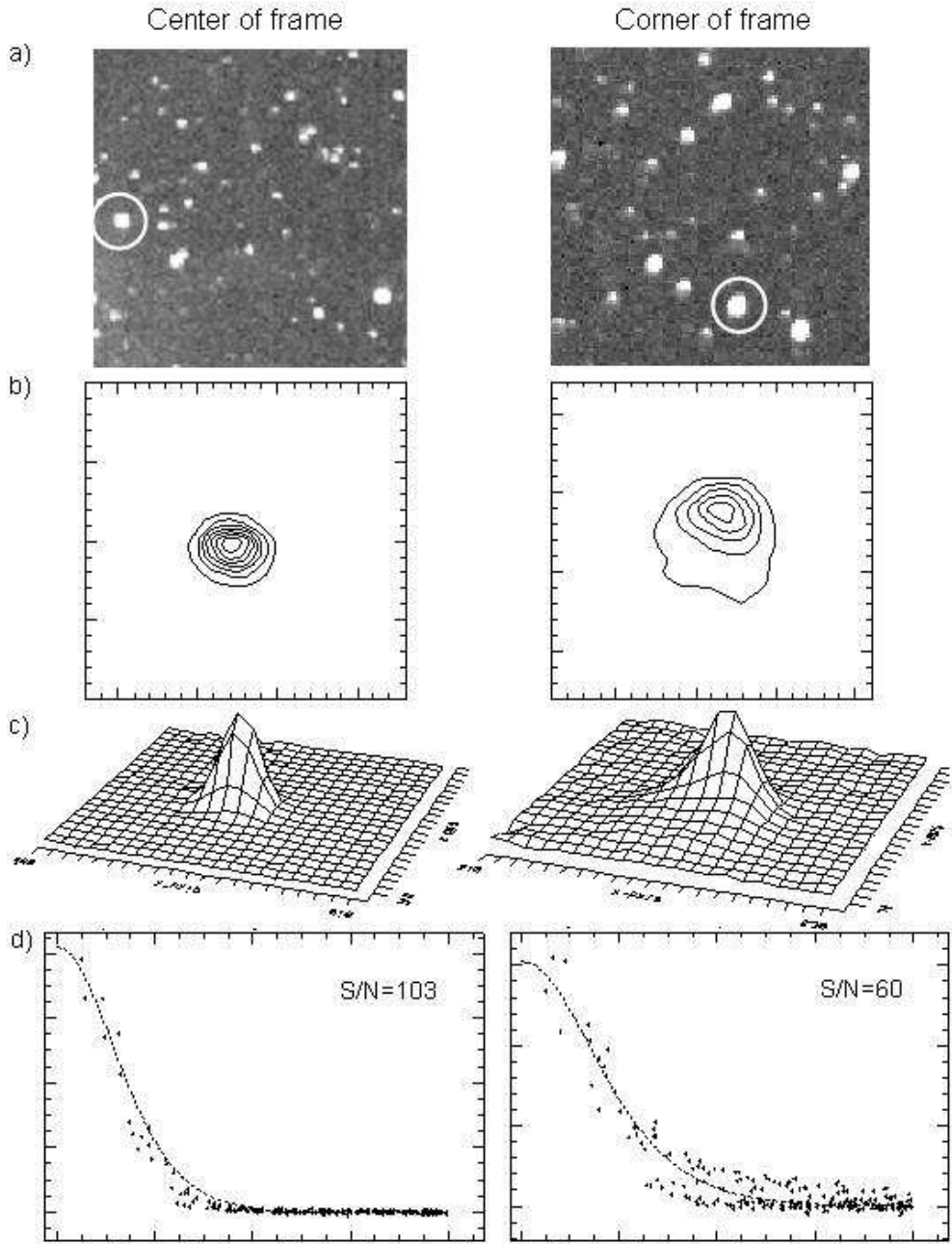}
\caption{Illustration of the effects of aberrations in our images. This figure shows contour (b), surface (c), and radial profile (d) diagrams of two stars of magnitude I=11.0 located respectively at the center and the corner of one of the survey frames . In (a) we show portions of the frame containing each star. The stars at the edges of the images are strongly affected by aberration. Both stars have a FWHM of about 9 arcsec (the pixel scale in the images, after binning 2$\times$2, is $\simeq$4.8 arcsec/pix), but the S/N of the aberrated star is 1.7 times worse than the S/N of the star at the center of the image.\label{fig4}}
\end{figure}

The telescope-camera combination was selected as a trade between aperture, sky coverage, and pixel scale. We needed the largest practical field of view that a commercially available CCD camera could cover. This translated into a demand for short focal length, which we met by choosing an 8-in Meade telescope with an optional f/6.3 focal ratio. The usable field of view of this telescope without vignetting is 0.37$\times$0.37 degrees, but we are using the entire field provided by the CCD ($\simeq$ 1.28$\times$1.28 degrees) which, in spite of the vignetting and coma aberration in the images, still provides acceptable photometry. The coma results in an increasing radial elongation of the shape of the stars as we move away from the center of the images. This aberration, combined with the vignetting, results in a progressive decay of the S/N of the stars toward the edge of the images. These effects are illustrated in figure~\ref{fig4}, where we show how the shape, and the S/N of two stars of similar magnitude vary depending on the location of the stars in the images.

The CCD is the largest available from Apogee Instruments Inc. (a 28.7$\times$28.7 mm chip, with 0.014 $\mu$m pixels). On the Meade telescope this yields a pixel scale of 2.256 arcsec per pixel. In practice the seeing at our site is frequently so bad that we bin 2$\times$2 to reduce the readout time. Thus an after-the-fact analysis shows that we could have saved about 6,000 dollars by choosing a 1024$\times$1024 CCD with 24$\mu$m pixels.

To reach our magnitude limit (I $\sim$ 15) with a S/N of 3 requires 3 minute integrations\footnote{This statement applies to the central 1.0 x 1.0 degrees of each image. For stars closer to the edges the S/N ratio worsens because of the aberrations}. At that integration time, stars brighter than I $\simeq$ 8.5-9.0 saturate. A larger telescope with smaller field would accomplish this faster, but CCD readout (10-15 seconds in the non-binned configuration) and slewing between fields ($\simeq$ 20 seconds) would diminish the duty cycle, resulting in little gain for the extra cost. In three minute exposures, the noise from sky brightness in the I-band dominates the dark noise from the camera, which is 2-stage thermo-electrically cooled to -25 $\pm$ 1-2 $^\circ$C. The CCD is a front illuminated Thomson THX7899M device. Its dark current at operating temperature is $\sim$ 0.16 ADU/sec. The CCD gain is $\sim$ 14 $e^{-}$/ADU, and the read-out noise is $\sim$ 30 $e^{-}$. In the 2$\times$2 binned mode, readout requires 5-6 seconds.

The telescope was delivered with a Meade LX-200 computerized mount, which we used to begin operations. We quickly realized that this mount was inadequate in 2 respects. The pointing accuracy, even after applying careful corrections, was no better than 8-10 arcminutes, and tracking trails 10-15 arcseconds long appeared after 3-4 minute exposures. These trails and pointing errors made the automated astrometry (see section 4) very difficult and time consuming. In addition, the mount began to experience random runaways in RA after a few month's operations. We decided to replace this mount with a much better quality (and more expensive) one, the german-equatorial GT-1100s Paramount from Software Bisque. The pointing accuracy with this new mount is better than 30-45 arcsec after correcting systematic pointing errors using a TPoint model\footnote{The TPoint software package is included the Professional Astronomy Software Suite distribution from Software Bisque}. The slew rate of the Paramount, 5 deg/sec, and the large field of view of our telescope, also makes the system ideal for rapid response to targets of opportunity, such as gamma ray bursts\footnote{See \citet{lop02} and \citet{nys03} for a sample of the Pisgah Survey contributions to this field}.

The camera is mounted to the telescope using a custom made connector that includes room for a single 50mm filter. Our intended targets are very red, so we have inserted a Bessell I-band filter, with a central wavelength of $\lambda_{center}$ = 8000 $\AA$, bandwidth $\Delta\lambda_{FWHM}$ $\simeq$ 1500 $\AA$, and a peak transmission of $\simeq$ 97 $\%$. Figures 5 and 6 show the final instrumental setup. Total hardware costs, including the control software, appear in Table~\ref{tbl-2}.

The telescope and the CCD are controlled by programs from Software Bisque called TheSky and CCDSoft, which are in common use among amateur astronomers. We have wrapped these commercial products inside a set of custom observing routines that coordinate the operations of all the hardware, including the dome. The result is a fully automated system for data collection. We describe the software in more detail in section 4.1.

\subsection{Location and Building Enclosure} 
 
The selection of the site to install the survey's equipment was a compromise between sky quality\footnote{The I-band sky brightness at PARI has been measured using 12 images at random, collected by the Pisgah Survey from August trough November 2000. All the images were taken around midnight, at an average zenith distance of 45 degrees. The results are shown in Table~\ref{tbl-3}} and proximity to Chapel Hill. Since this was to be a prototype, and likely to require a lot of troubleshooting, we decided to accept an offer from the Pisgah Astronomical Research Institute (PARI)\footnote{http://www.pari.edu} to install the hardware at their facilities in Rosman, North Carolina, 270 miles away from our home institution. PARI also provided site development and constructed the enclosure at their expense.

PARI personnel constructed the building to house the telescope at the highest point in the site, with an elevation of about 1,800 feet above the sea level. The building consists of a two-story structure that includes battery backup power, air and humidity evacuation systems, a control computer, and fiber optics connections to allow remote access to the equipment. The structure is topped by a 10-foot diameter dome acquired from Technical Innovations Inc. The dome includes the Digital Domeworks software package, which provides the necessary tools to control the dome, and a small weather station to ensure automatic closure in the event of adverse weather conditions. Besides the weather, which is unusable 40$\%$ of the time, the dome has been responsible for most of the lost of observing time, suffering a variety of mechanical and electrical failures. Future projects at PARI will use roll-off roofs for improved simplicity and reliability.
 
\subsection{Computers, GPS, and remote access}

Survey operations and subsequent data analysis are handled by three fully dedicated computers. The first one, located in the first floor of the survey's building at PARI, is a pentium III, 550 MHz PC that runs the survey software and controls the telescope, camera, and dome. It also serves as temporary storage for the data collected each night. The second computer is a pentium II laptop that provides remote access to the survey at any time, from any location with ethernet or telephone, using the software package PCAnywhere from Symantec Corporation. Ethernet access to PARI is via a T1 line, which allows us to transfer the approximately 0.4 Gb of data collected each night at a speed of $\sim$ 0.24 Mb/sec. The third and last computer is a Dell Precision Workstation 530 that is used to reduce and analyze the data.

In addition to these three computers, the survey accesses two other machines in the PARI local network that display, respectively, the local weather conditions through a set of TV cameras placed around the site, and information from an on-site GPS receiver. The internal clock on the survey computer synchronizes itself to the GPS each 20 minutes, to assure precise tracking of time in our time series photometry.

\section{Data Collection and Analysis Routines}

\subsection{The Automated Observations Routine}

The most difficult cost to contain in an automated telescope project is software development. It is in this category that we have realized the greatest savings by exploiting consumer-level software. Each one of the commercial software packages used to operate the telescope, the CCD, and the dome include libraries of Active X functions developed by the Astronomy Common Object Model Initiative\footnote{http://ascom-standards.org/index.html} (ASCOM). Those functions can be called from Visual Basic or other Microsoft languages. We have written a Visual Basic script that manages the automation of the telescope, dome, and CCD through these library routines. A flow chart of the program is shown in figure 7\footnote{The entire source code is available from the authors.}. We are currently in the process of porting the code to C$\#$ in the $.NET$ framework, which will allow automatic response to web generated interrupts for observing Gamma Ray Bursts afterglows and other targets of opportunity.

The program, called {\it SurveyLoop.vbs}, runs continuously as a background process in the computer installed at the dome. It keeps track of the time of the day through the computer's internal clock. Two hours before sunset, it begins to cool the CCD to the preset temperature of -$25^{\circ}$C. At sunset, after checking the weather conditions, the program opens the dome and slews the telescope to a position close to zenith, where it begins taking a series of sky flats, followed by dark frames. Once it finishes taking the darks, the program waits until the end of the astronomical twilight to begin monitoring the list of preset target fields. The integration time for all the fields is 180 sec, which yields a S/N $\simeq$ 3 for I= 15.0.

Initially our list of targets were located along the celestial equator, to allow follow-up observations from either hemisphere. The telescope re-pointed in RA after each exposure, monitoring those fields on the list of targets that fell within $\pm$ 2 hours of HA = 0. Using this strategy we were able to repeatedly image 60-70 different fields per night, covering an sky area of 107.0 - 115.3 sq. deg. During the first year of operations, we found two major problems with this scheme. First, the fields were imaged only 4 times per night, and we discovered through experience that this made it impossible to collect enough fields for binary detection in a single season. Second, the frequent long slews in  R.A.  required the already problematic dome to move quite often, leading to frequent misalignment problems between the shutter and the telescope, and to mechanical failures. To avoid those two problems, we changed the spatial distribution of target fields to sets of 10 (2x5) adjacent fields	in a rectangular area of $\sim$ 16.5 sq. deg. (2.57 x 6.42 degrees). The survey cycles over the fields in one set while they are within hour angles of $\pm$ 3 hours, then moves to the next set of fields, strategically located 5-6 hours later in RA. Each set of fields is monitored until $\sim$ 300 clear images are available, or until it falls out of the HA coverage\footnote{given the site's characteristic weather quality, it usually takes 40-60 nights to collect at least 300 good frames}. We avoid fields in the plane of the Galaxy, because they are too crowded for automated reduction to work properly (see section 4.2.1). This is not a significant loss, because our survey will not cover the whole sky and the distribution of our target stars is essentially isotropic (our magnitude limit defines a volume that does not extend out of the galactic disk for M-dwarfs). Our new strategy has reduced the number of dome failures dramatically, and allows us to acquire at least 8-10 images of each target field per night, with a cycle time of about 35 minutes. So it is in principle possible to collect enough frames for binary detection in one observing season.

Between each exposure, the program checks the weather conditions and the time. If either an {\it adverse weather} or {\it morning twilight} notice occurs, the equipment is shut down following one of the two procedures outlined in figure~\ref{fig7}. If the notice indicates adverse weather conditions, the program closes the dome, parks the telescope, and remains idle, ready to resume observations if the weather conditions improve. If the notice reports the beginning of the morning astronomical twilight, the program parks the telescope, takes a second set of dark frames, and turns off the CCD cooler. It then goes back to the beginning of the loop, where it waits until the next sunset.

\subsection{The Reduction and Analysis Routines}

The survey collects a total of 130-200 images per night, including flats, darks, and target fields. These are analyzed off-line by the reduction and analysis protocol outlined in figure 8. The reductions are done at the end of each night, but the analysis to find eclipsing variable candidates must include more than a single night's data. Our current experience shows that at least 10 observations during eclipse are required to trigger our period search routine (see 4.2.2). If we assume the observations occur at random orbital phase, then 300 observations are required to insure a greater than 95$\%$ detection probablilty for binaries with eclipses equal to 5$\%$ of their orbital period. The exact number of observations depends on the ratio of the 
	     stars's radii, which we do not know a-priori, and on the 
	     photometric precision of the light curves. For binaries with longer eclipses, the number of observations needed is smaller than for those with shorter eclipses. In addition, the number of observations necessary to detect the eclipses is larger for fainter stars, as the photometric dispersion of the light curves increases with magnitude. As a result, the survey will be increasingly incomplete because we remove fields from the target list after 300 frames are collected\footnote{Our survey will also be incomplete for binaries with orbital periods near integer multiples and fractions of 1 day, since our observations are not really random in time.  We reduce this effect by observing the same fields in successive months, but cannot entirely eliminate it from a single site}. A full analysis following the scheme in figure 8 includes all the images collected for each field, which usually take several months to acquire.
 
Our technique for finding variables relies upon having ``light curves'' of each star available, so the analysis must automatically identify stars based on their positions and assemble photometric measurements from multiple images into a time series. Only then can we look for the periodic brightness changes that will signal the discovery of an eclipsing binary, or any other kind of periodic variable.

In our reduction and analysis routines, we have again saved on software development by relying heavily on available code, mostly within the ``Image Reduction and Analysis Facility'' (IRAF) packages from NOAO. Our protocol consists of three main programs, {\it genera$\_$redroutine.C}, {\it genera$\_$analroutine.C}, and {\it period$\_$search.C}. These programs generate IRAF scripts called {\it reduce$\_$all.cl}, {\it analyze$\_$all.cl} and {\it search$\_$period.cl}. These scripts then direct a fully automated analysis of the data within IRAF. The final output is a set of folded light curves of all the variable star candidates detected in the monitored fields. We describe in the following subsections each one of the scripts in more detail.

\subsubsection{reduce$\_$all.cl}

This script reduces the images of the target fields collected each night, and performs aperture photometry in all the stars contained in those fields. Its inputs are ``masterdark'' and ``masterflat'' images and the list of target fields monitored that night. Its main outputs are the reduced images, photometry files containing the pixel coordinates of all the stars in each field, and instrumental I-band magnitudes of these stars. 

The generation of the master dark and master flat is the only non-automatic part of the reduction, since each calibration frame used to generate these master images must be inspected by hand to avoid introducing spurious effects in the reduced data. Master darks are generated by taking the average of all the dark frames collected in one night; the master flats are constructed in a similar manner, but applying an additional {\it $\sigma$$-$clipping} algorithm to each flat, in order to eliminate any visible stars. Whenever calibration frames are not available for a given night, the most recent ones from previous nights are used instead.

The script first performs standard IRAF CCD image reductions (see Massey 1997), and then rotates the images to correct for the known offset between our frames and the orientation the POSS2 Digital Sky Survey plates (see next subsection), and renames them in preparation for automated astrometry. As shown in the flow chart (figure 8), the script then runs the Sextractor routine of \citet{ber96} to identify stellar images in the frames. In comparison to similar routines, we have found this program provides the best results in identifying stellar objects in our images. The script eliminates completely cloudy frames by throwing away images with fewer than 200 stars. In addition, images with more than 2000 stars are set aside as crowded fields, for which our automated aperture photometry yields poor results, as stellar blends begin to proliferate given our low pixel resolution. We have failed in various attempts to solve this problem by implementing the IRAF $Point$ $Spread$ $Function$ (PSF) photometry algorithms because, as a consequence of the coma aberration present in the images, the PSF of the stars varies with position in the frames, and can't be modelled easily. At present we have opted for restricting our analysis to aperture photometry, but we are considering the addition of a variable PSF photometry routine to the survey's analysis software. These fields cataloged as {\it crowded} are removed from the target lists.

On the images that remain, the script performs aperture photometry (using the IRAF task ``phot'') in two passes. The first pass uses the same aperture size for all the stars to obtain an estimation of their instrumental magnitudes; then the second pass optimizes the aperture sizes for each star based on those magnitude estimations. Finally, the images and their associated photometry lists are sorted into directories according to their R.A. and Dec.

\subsubsection{analyze$\_$all.cl and search$\_$period.cl}

The outputs from {\it reduce$\_$all.cl} are the inputs to the next script {\it analyze$\_$all.cl}, which derives astrometric solutions for each image, and cross-identifies each star in all the images of a given field so light curves can be generated. The script uses the package WCSTools \citep[2001]{min97} to perform astrometric solutions for each target field, using list of stars from the corresponding images of the Digitized Sky Survey POSS2 as references. After finishing the astrometry, the script can make light curves from the dates and times in the image headers, and the magnitudes from the photometry files, but the latter must first be corrected for extinction and calibrated against a standard system.

To accomplish this, we select a reference frame for each target field from the night with the best photometric quality. The selection of the reference frames need to be done by hand, although only once per target field. For the reference frames only, we conduct calibrated photometry to produce a single reference image of each survey field. The raw magnitudes in that frame are transformed to the Johnson-Cousin system by using the magnitude calibration technique by \citet{har81}. We use $l = 3 - 4$ images of each target field, from the selected night, with airmasses $\chi \in [1,2]$. In each image we select as standards $m = 10 -12$ isolated stars, uniformly distributed across the frame (avoiding the edges). The selected objects have no saturated pixels, and have magnitudes published in the Tycho-2 Catalog \citep{hog00}, with Tycho $B_{T}$-$V_{T}$ colors in the range -0.2 $\leq$ $B_{T}$ - $V_{T}$ $\leq$ 2.0. The Tycho standard magnitudes of those stars are converted to I-band Johnson-Cousin magnitudes using the transformation equation derived by \citet{ric00}. Finally, the raw magnitudes are transformed using the standard transformation equation 
\begin{equation}
		I_{raw} - I_{std} = a_{1} + a_{2}\cdot\chi 
	\end{equation}
This equation includes no color terms because we monitor the stars in only one filter, so no color information is available. The calculated values of the $a_{1}$ and $a_{2}$ coefficients are $a_{1} = 9.509 \pm 0.097$ and $a_{2} = 0.051 \pm 0.070$. The  uncertainties in the estimated $I_{std}$ magnitudes are of the order of $\sigma_{I} \simeq 0.13$ mags.  

The raw magnitudes of all the other images of each field are corrected to the level of the reference frame by calculating and applying a single magnitude correction $\Delta$I.  This correction is an average of the difference measured for each star in the frame.  Magnitude difference diagrams for some sample images are represented in figure 9, and show that there is no systematic bias with brightness in the photometry.  We discard images with $\Delta$I $\geq$ 0.5, because their photometric quality is too poor.

Once the magnitudes are adjusted in this way, the script assembles a separate light curve for each star, including now {\it all} the data collected up to this time. It is possible to identify variables from their light curves at this time using the strategy employed by \citet{poj97}, who selected stars with standard deviations $\sigma_{\overline{I}}$ larger than the expected for their $\overline{I}$ magnitude. Figure~\ref{fig10} shows graphically how this strategy would work for one of the fields monitored by the survey; stars above the dashed-line in the figure are variable candidates. However, this strategy is not optimum for our survey, because it does not use all the available information. We know that eclipsing binaries we seek are are {\it periodic}, and can implement more sensitive techniques by searching for periodic variations. We have chosen to use the technique of ``analysis of variance'' (AoV) \citep{sch89}, which was also employed as a period search method by ASAS \citep{poj97}, and which excels at detecting narrow periodic features.

The plot in Figure~\ref{fig10} is from a field that includes one of the three known eclipsing M-dwarf binaries, GJ 2069A. As we discuss in section 5, this binary has eclipses only $\sim$ 0.2 magnitudes in depth. Its location in figure 10 is marked with a dark square, and does not clearly stand out compared to the other variable star candidates. Many of the stars that fall above the curve are false positives caused by blended stellar profiles. If we apply a selection threshold capable of extracting GJ 2069A from this diagram, it will generate too many false candidates to be useful. On the other hand, the AoV method finds GJ 2069A easily by looking for variance in its folded light curves, as demonstrated in section 5.1.

The last script in our analysis performs the AoV method on all of the light curves generated for each field observed. We have used the public {\it aov.f} routine written by \citet{sch89}, but we call it from IRAF for convenience in automation. The program applies a $One$ $Way$ $Analysis$ $of$ $Variance$ algorithm, which \citet{sch89} has shown is much better than the Fourier transforms at detecting periodic narrow pulses. After folding the light curves over a specified range of trial frequencies, the program generates lists of values of the standard AoV test statistics $\Theta_{AoV}$ for those frequencies (we will call plots of $\Theta_{AoV}$ vs. frequency {\it ``AoV periodograms''}). Any periodic feature in the light curves appears in those AoV periodograms as a high peak at the corresponding frequency. We refer the reader to the original paper by Schwarzenberg-Czerny for a more detailed description of this technique. For our purposes, we have restricted the range of trial frequencies to 0.1 - 10 $days^{-1}$, since, as indicated in section 2.1, the probability of detecting binaries with orbital periods larger than 10 days is close to zero.

To identify candidates from the output of the AoV routine, we define a 
"contrast parameter" by calculating the ratio of the largest peak above 5 
$days^{-1}$ and the largest below 5 $days^{-1}$. When that ratio is less than a given threshold value the star is considered a variable 
candidate. The value of the threshold was calibrated using the AoV periodogram of a variety of variables, the M-dwarf binary GJ 2069A among them (see section 5.1). This new identification technique eliminates the problem of false detections caused by blends, which do not show, in general, any detectable periodic behavior.

Finally, the script automatically generates, for all the possible variable stars, {\it supermongo} \citep{lup98} plots of their light curves folded at the periods identified in the AoV routine. The folded light curves must be inspected by eye to make a final determination about the existence and nature of the variability.

\section{Preliminary Results}

In this section we present the results of the analysis of the first 9 fields monitored by the survey. These fields cover an area of sky of 14.76 sq. degrees. Their central coordinates and the number of times that they have been monitored are listed in Table~\ref{tbl-3}. Also listed in that table are the number of stars per field, and the number of variables found in each one of them. The area surveyed contains a total of 8,201 stars brighter than I = 15. This is after eliminating from the stellar census all the objects that were detected in fewer than 50 of the frames, which we regard as artifacts. 

Our analysis software currently selects from these fields 671 stars as possible periodic variables, i.e. about 8.2$\%$ of all the stars. After subsequent visual inspection of the folded light curves, this number is reduced to 20 definite periodic variables, the faintest of which is I = 13.4. Visual inspection is a very time-consuming way to make this final cut, so we are now in the process of developing and training selection algorithms that operate on the folded light curves.  With a larger number of fields, we will also be able to better adjust the selection threshold for the AoV periodogram. Once these two selection steps are optimized, we expect to minimize  the number of candidates that require final inspection by eye.

Table~\ref{tbl-4} summarizes the main parameters of the 20 variables identified. These parameters are the calculated R.A. and Dec for those stars, their average apparent magnitude $\, ={I}$, their estimated magnitude variation amplitude $\Delta I$, their period $P$, their estimated epoch of minimum brightness $T_{0}$, their variability classification, and the results of an extensive cross-identification search among all the available on-line variable star catalogs. Only 5 of those objects have been previously cataloged as variables, or suspected variables. The rest are new discoveries. Among the detected variables are 6 detached eclipsing (algol type) binaries, 2 contact (W Uma) binaries, 1 cepheid, 1 RRc Lyrae, 1 semi-regular variable, 2 ``long-term'' variables, and 7 variables of unknown type\footnote{These variable candidates present very clear peaks in their AoV periodograms, but we haven't been able to classify them from the features in their folded light curves}. The two objects classified as ``long-term'' variables show clear magnitude variation trends, but not apparent periodicity. If periodic at all, their periods must be longer than our current follow up time of 41 days. Figure~\ref{fig11} shows the light curves of the 11 periodic variables of identifiable type. On the left column we show their light curves in MHJD time space; the curves on the right are the same light curves after folding them with the estimated periods. We also show in figure~\ref{fig12} the light curves of the two ``long-term'' variables. For PS-3-vs0048 we have included available data from The Amateur Sky Survey (TASS) \citep{ric00}. The TASS data extends the photometric coverage of this object by 20 more days, and strengthens our long-term variability hypothesis.

In addition to the variables that we detected, there are three objects previously cataloged as variables in these fields that we did not detect. They are listed in Table~\ref{tbl-5}. The first two objects have been cataloged as suspected variables; the third one is classified by the $Combined$ $General$ $Catalogue$ $of$ $Variable$ $Stars$ \citep{kho98} as an ``unstudied variable with rapid light changes''. It is possible that those light changes are faster than our 180 $sec$ exposure time, or their amplitude is much smaller than the magnitude dispersion of our light curves, or that they are aperiodic.  In any case, we are not troubled by the non-detections, since their light curves appear very unlike the eclipsing binaries we seek.

\subsection{GJ 2069A}

In order to test the performance of our variable identification technique, we included the previously known M-dwarf binary GJ 2069A among the first set of stars monitored by the survey. This binary provides in fact an excellent test case, given that its characteristics are far from optimum for a successful detection in our images, as described below. Admirably, our routines have no problem identifying the object as a variable candidate (see figure~\ref{fig15}), although some problems were found when we tried to recover its orbital period from the folded light curves. 

The orbital inclination of GJ 2069A is very high, which translates into shallow eclipses, of only 0.2 and 0.15 mags in depth \citep{del99}. These values are close to the dispersion limit of our photometry for the apparent magnitude of the binary. Therefore its light curve is at the marginal limit of the survey for a successful identification by visual inspection. Furthermore, given the low resolution of our images, the binary appears blended with a brighter nearby star, as shown in figure~\ref{fig13}. The magnitude of the blend is I $\sim$ 9.02. As we mentioned in section 3, stars of that magnitude saturate in some of the images, resulting in a larger magnitude dispersion in their light curves. In addition, both stars in the blend are M-dwarfs, and therefore magnetically active, with sporadic flares and continuous surface brightness variations. This larger magnitude dispersion, combined with the shallowness of the eclipses, inhibits a clear identification of the binary from its folded light curves (notice that we list it in Table~\ref{tbl-4} as a variable of type "unknown", because we could not recognize its Algol type light curve in our visual inspections). Figure~\ref{fig14} shows the current light curve of the blend in time space (top), and after folding it using the known orbital period of 2.771468 days, derived by \citet{del99} (bottom). The secondary eclipse (at phase $\sim$ 0.5) is obvious in the folded light curve, but only because we know a priori its location. There is no data during primary eclipse, which would occur around phase = 0 in that plot. The magnitude dispersion of the baseline is significant, because of the combination of effects mentioned above. 

With more data covering all the orbital phases, even a binary in such marginal conditions in our survey as GJ2069A can be eventually identified by eye. But with the current light curve coverage, the binary is already clearly identified as a variable star by the AoV periodograms, as illustrated in figure~\ref{fig15}. This figure shows the combined AoV periodogram of GJ 2069A and its blended companion (top), compared to the AoV periodogram of a non-variable star (bottom). While the periodogram of the non-variable is basically featureless, several variance peaks are clearly visible in the one for GJ 2069A. In particular, the peak at a frequency of about 0.361 $days^{-1}$ in the periodogram reproduces closely the mknow orbital period of $P$ $\sim$ 2.771468 days for the binary, proving the viability of our variable candidates selection algorithms

The identification of GJ 2069A as a periodic variable demonstrates that the Pisgah Survey's search strategy can successfully detect M-dwarf eclipsing binaries. However, our less successful identification of its period or light curve type shows that it will not always be possible to go from detection to follow-up observations without an intermediate step. Better seeing conditions and a slightly higher pixel resolution can reduce the number of cases where blends of this type become a problem. Also, continued Pisgah survey measurements of the fields with periodic variables of unknown type can ultimately resolve their light curves, as more cycles are measured, but this is not the most efficient use of survey time.  More practical in this case will be follow-up photometry from other automated telescopes, such as those designed for Gamma Ray Burst optical transient observations, which usually have ample time between bursts to conduct synoptic observations.

Even without intermediate observations, two thirds of the periodic variables that we detected have identifiable light curves, and can be followed up immediately. At this point, none of these are the low mass eclipsing binaries we seek, but this is consistent with our expected detection rate of 1 every $\sim$ 30 sq. degrees.
  
\section{Conclusions}

We have presented a technical description of the Pisgah Survey, a project intended to discover new low-mass detached eclipsing binary stars.  These new binaries will be used to add precision measurements to the mass-radius relation of low-mass main sequence stars, an important and timely task that will provide the observational data necessary to test the most current low-mass stellar models.

The Pisgah Survey demonstrates that it is possible to address astrophysically interesting projects within a modest budget by using equipment from the consumer astronomy market. Although this is not a new approach, as shown by previous successful projects such as ROTSE \citep{ake00} and STARE \citep{bro00}, none of those projects has recognized the material and labor savings of the Pisgah Survey: all our hardware, automated observing routines, and data analysis pipelines were assembled by one graduate student and staff personnel at PARI with a budget of about $\$$80,000. The greatest savings came in software, traditionally the most expensive and time-consuming part of small telescope projects.  The Pisgah survey shows that efficient automated observing routines and data analysis pipelines can be built from commonly available software with minimal additional programming.

Our preliminary results show that the survey can accomplish the task for which it was designed and also observe other astrophysical phenomena of interest, such as optical counterparts of Gamma Ray Bursts (see section 3.1).  Our variable search routines have successfully identified the previously known M-dwarf eclipsing binary GJ2069A from its AoV periodogram, although the object highlights a potentially important problem with identifying binaries in stellar blends, or when some of the pixels saturate (see section 5.1).  Furthermore, the survey has already discovered 15 new variable stars, and has re-identified another 5 previously known ones.  None of the new variables is clearly an M-dwarf, although one of them (PS-8-vs0031) is a detached eclipsing binary with a primary mass slightly over 1$M_{\sun}$.

In the future we will continue to observe selected fields, but modify our strategy to include the new nearby low-mass stars that are being found in the recently released 2MASS database \citep{cut03}. In this way we expect to increase our chances of success.  We also plan to optimize the variable identification algorithms and to implement automatic response to GRB localizations.

The greatest improvement we could make at this time would be to move the survey to a location with better weather conditions, but no funds are available for this purpose.  Given normal PARI weather, we expect to complete about 100 more sq. deg. in the coming year, which should yield at least 3 new low-mass eclipsing binaries, a number equal to the total number of systems studied to date.

\section{Acknowledgments}

We would like to thank UNC alumnus Dr. Henry Cox for his private donations to this project, and for suggesting the name ``Pisgah Survey''. We also thank Mr. Don Cline, president of PARI, for making the institute's facilities fully available to us, and for his contributions to the design and construction of the survey's building. We are specially grateful to all the PARI personnel for the incalculable amount of help that we have received from them on the setup and maintenance of our equipment. Special thanks to Dr. Eileen Friel, from the $National$ $Science$ $Foundation$, for revisions of this paper, and for her many constructive commentaries toward its completion, and to the referee Dr. Michael Richmond for his helpful comments and suggestions. This project has been partially supported by the National Science Foundation through grant AST 00-94289.





\clearpage


\clearpage

\clearpage

\clearpage

\clearpage
\begin{figure}
\epsscale{0.6}
\plotone{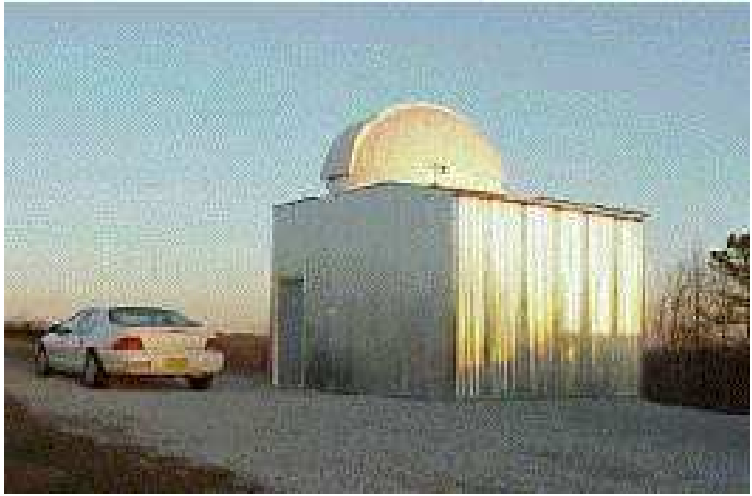}
\caption{Building constructed at the facilities of the Pisgah Astronomical Research Institute (PARI), in Rosman, North Carolina, to house the equipment of the Pisgah Survey. \label{fig5}}
\end{figure}

\clearpage
\begin{figure}
\epsscale{0.5}
\plotone{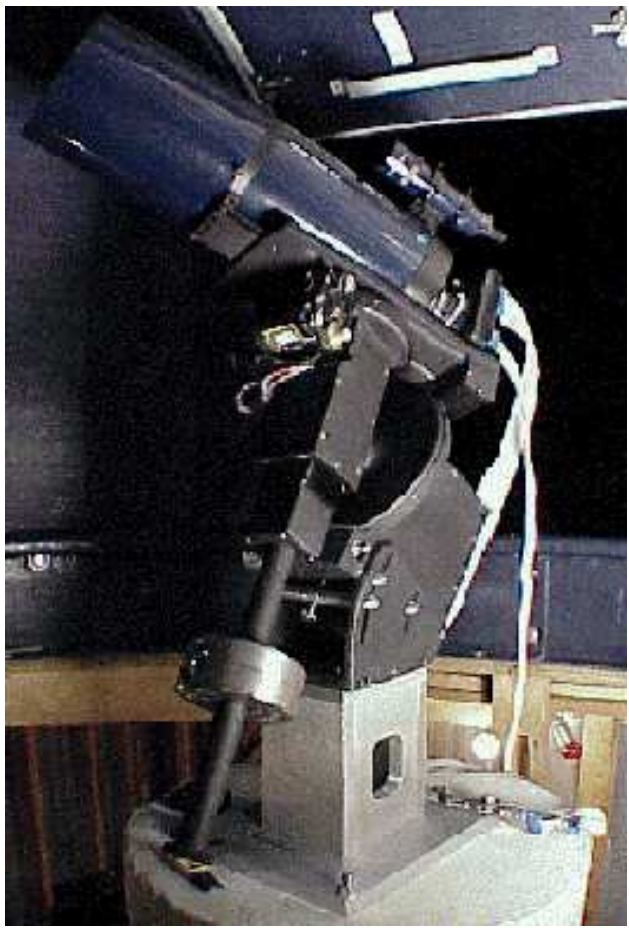}
\caption{Final setup of the survey's equipment. This includes the Meade 0.2-m telescope tube, the Apogee Ap10 CCD camera attached to the telescope through our custom made connector (inside of this connector is the I-band Bessell filter), and the GT-1100s Paramount from Software Bisque. \label{fig6}}
\end{figure}

\clearpage
\begin{figure}
\epsscale{1.0}
\plotone{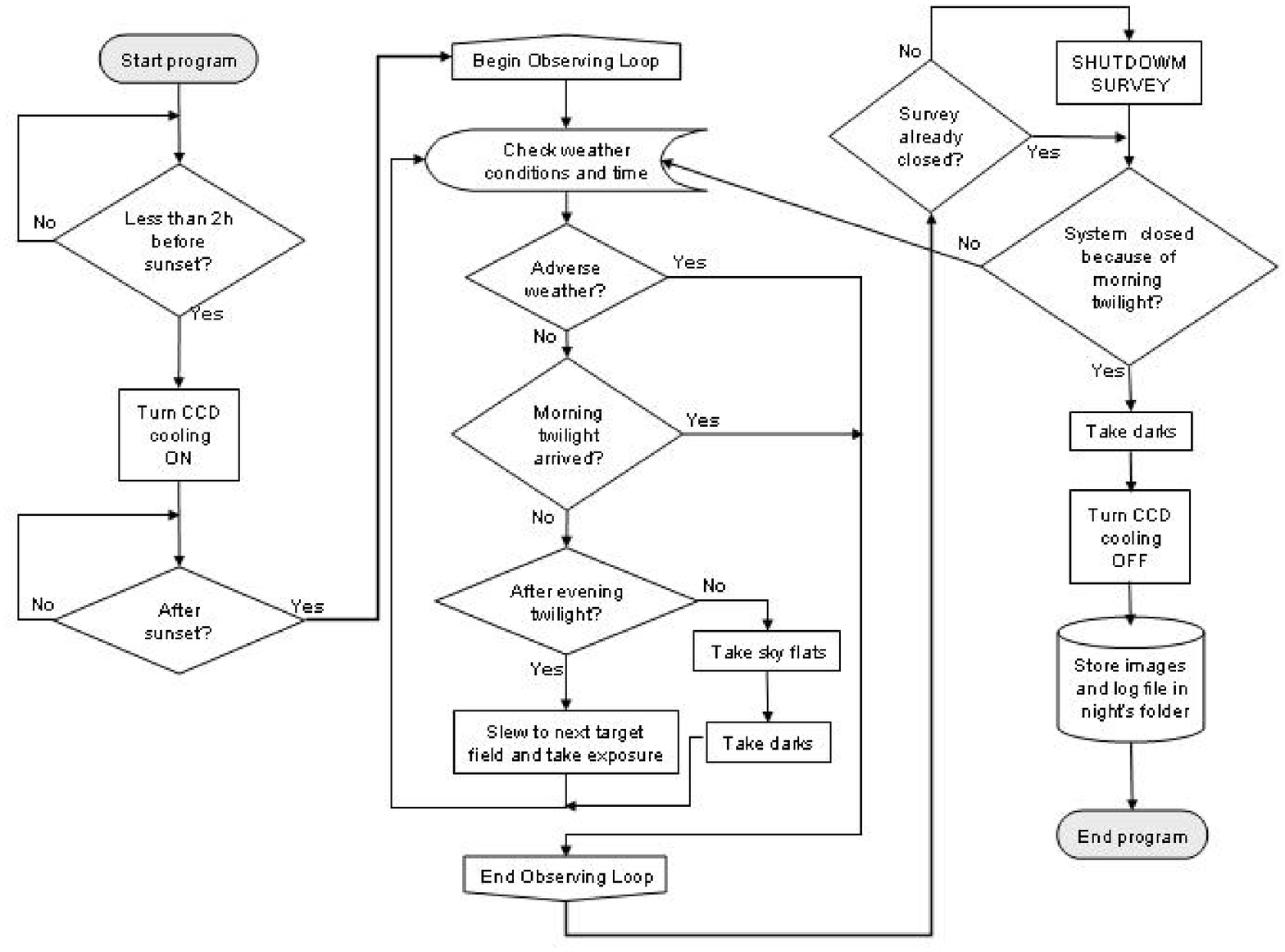}
\caption{Flow chart of the Visual Basic script {\it SurveyLoop.vbs}, which handles the nightly automated operations of the survey. \label{fig7}}
\end{figure}

\clearpage
\begin{figure}
\epsscale{1.0}
\plotone{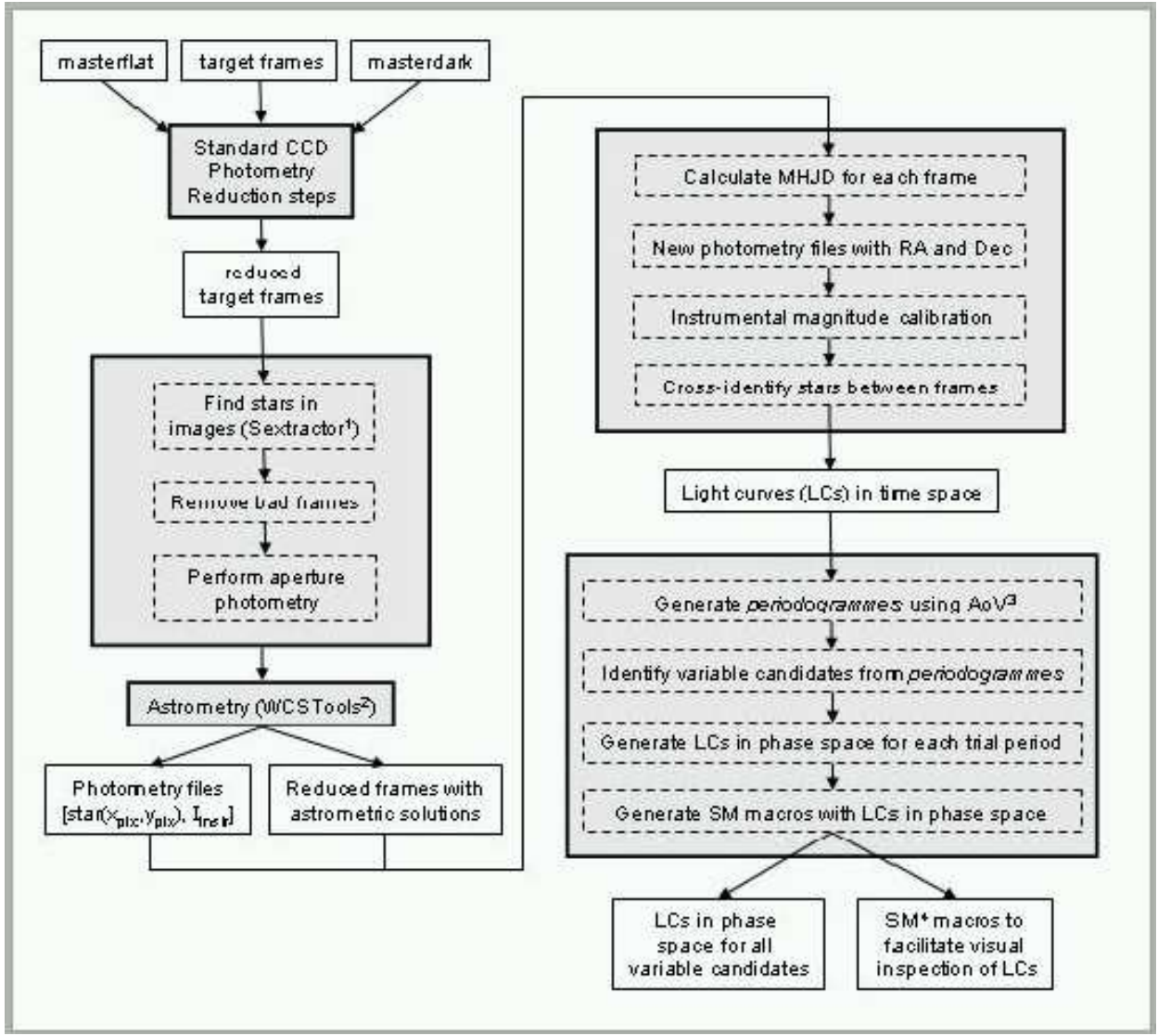}
\caption{Schematic outline of the data reduction and analysis protocol. A detailed description is provided in section 4.2.\label{fig8}}
\end{figure} 

\clearpage
\begin{figure}
\epsscale{1.0}
\plotone{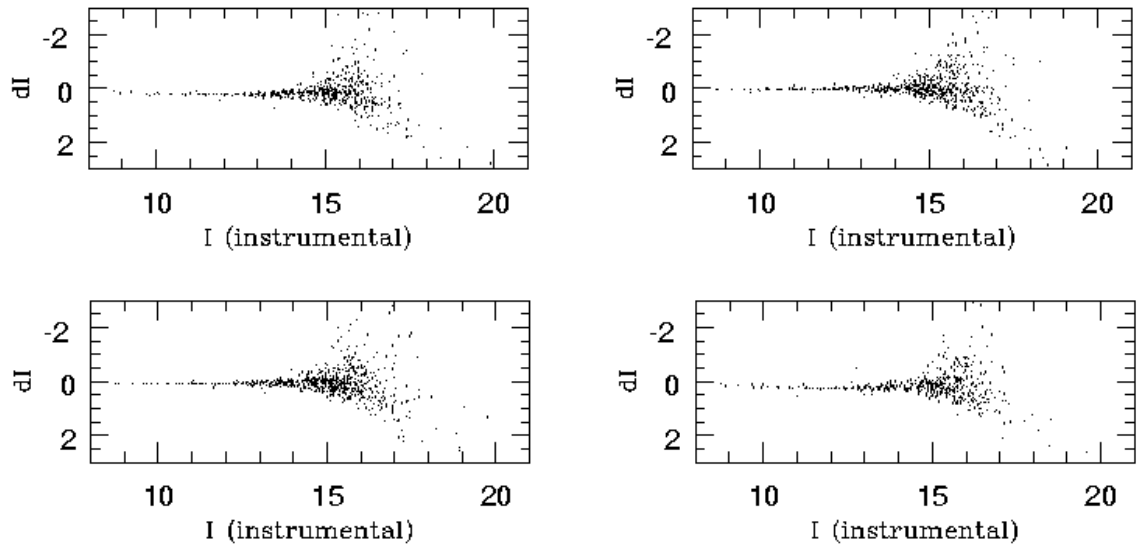}
\caption{Sample magnitude difference diagrams for four images of one target field, collected in different nights. The analysis routine measures the magnitude difference {\it dI} between the image and the reference frame, for each star. An average $\Delta$I correction is then applied to all the stars in that image. \label{fig9}}
\end{figure} 

\clearpage
\begin{figure}
\epsscale{1.0}
\plotone{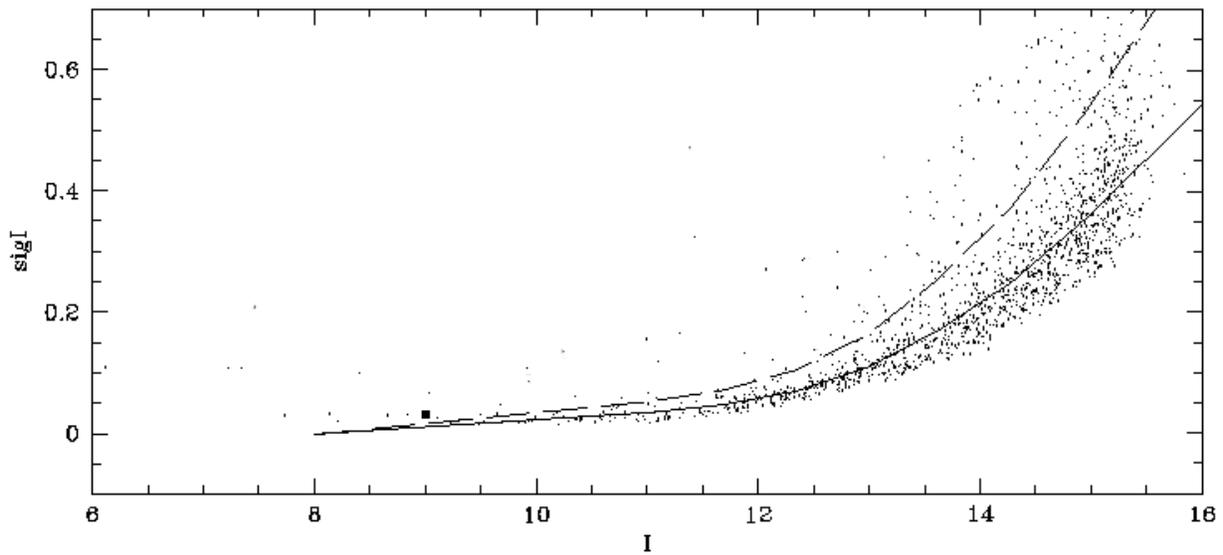}
\caption{Magnitude dispersion diagram for stars in the target field centered at RA = $08^{h}30^{m}$, Dec = $19^{\circ}24'$. The continuous line traces the average value of $\sigma_{\overline{I}}$ as a function of I. The line was computed by fitting an exponential function to the average $\sigma_{I}$'s of 0.5-mag bins between $I_{raw}$= 10.0 and $I_{raw}$= 15.0. The values of the curve at the bright and faint ends of the diagram were obtained by extrapolation of the exponential fit. The dashed line corresponds to 2-standard deviations above the average. The location of the M-dwarf detached eclipsing GJ 2069A in this diagram is indicated by the black square. \label{fig10}}
\end{figure}

\clearpage
\begin{figure}
\epsscale{1.0}
\plotone{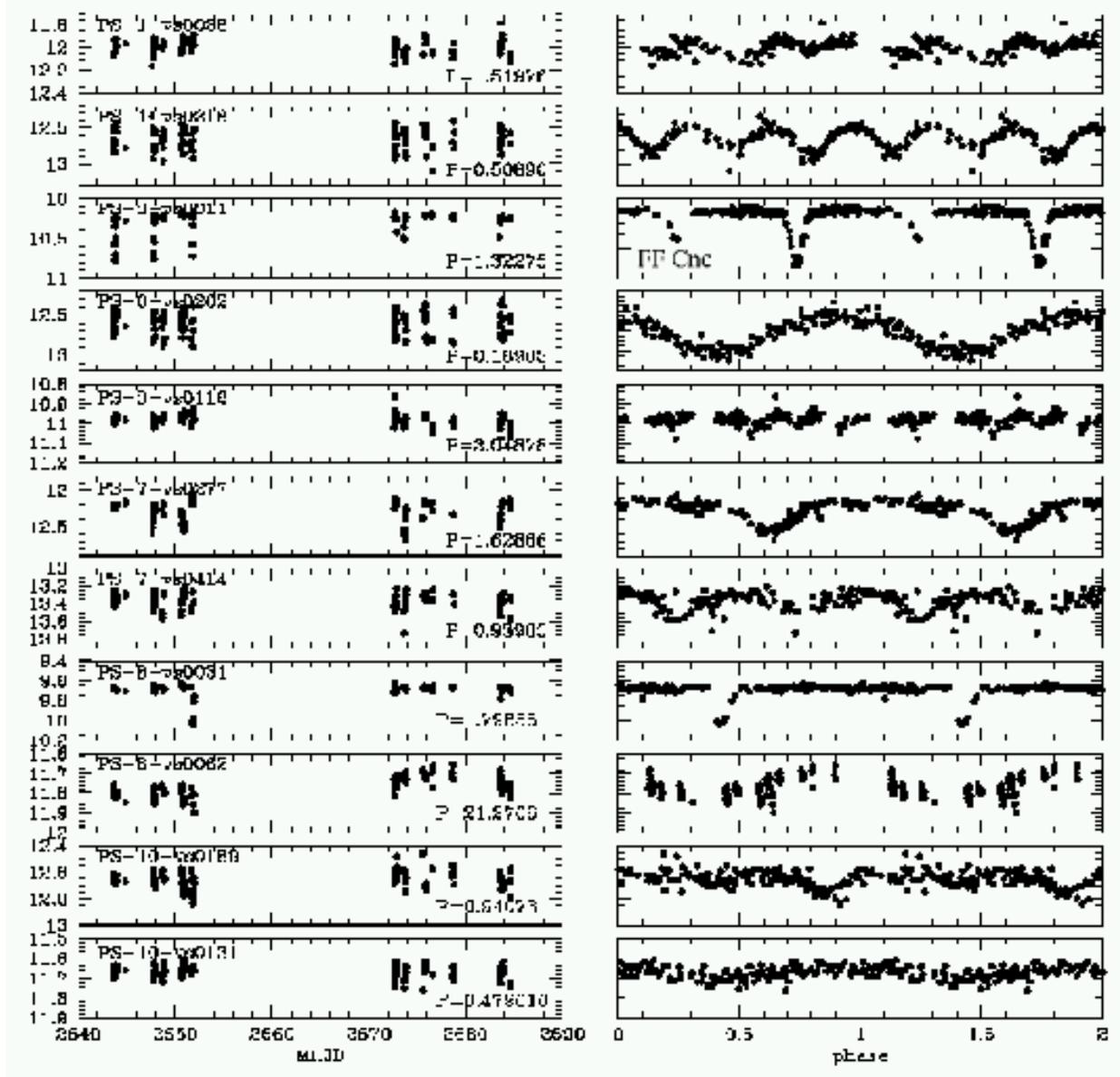}
\caption{ Light curves of the 11 confirmed periodic variables found in the first 14.76 sq. deg. of sky monitored by the survey. On the left we show the light curves of those variables in MHJD time space; the ones on the right are the same light curves after folding them with the estimated periods. \label{fig11}}
\end{figure}

\clearpage
\begin{figure}
\epsscale{1.0}
\plotone{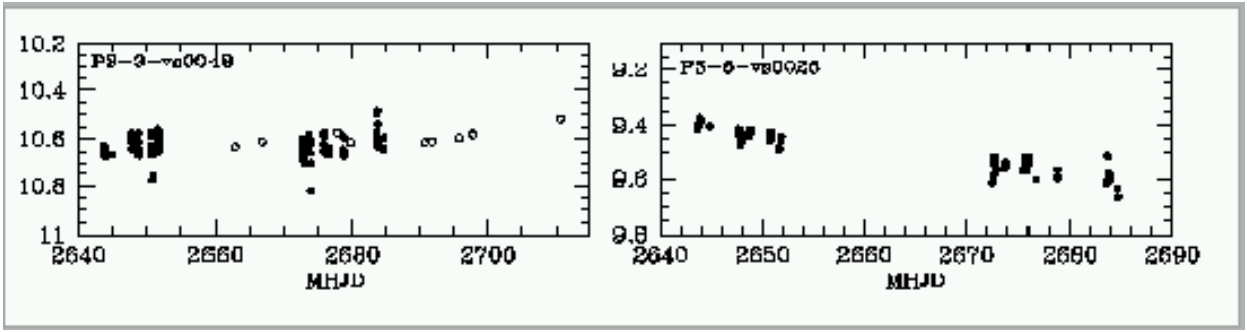}
\caption{ Light curves of the 2 suspected ``long-term'' variables discovered in  the first 14.76 sq. deg. of sky monitored by the survey. The open circles show the data from TASS. \label{fig12}}
\end{figure}

 \clearpage
\begin{figure}
\epsscale{1.0}
\plotone{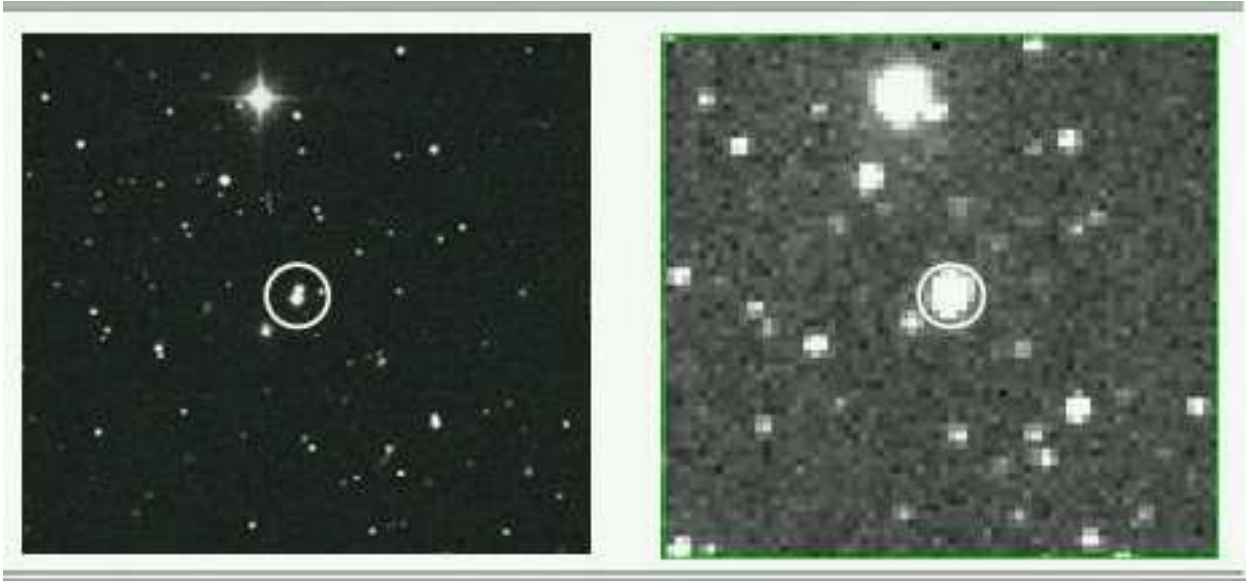}
\caption{Image of the M-dwarf eclipsing binary GJ2069A and its companion star (GJ2069B) from the Digitized Sky Survey POSS2 (right) and from the Pisgah Survey (left). Given the low resolution of the Pisgah Survey images (2.256 arcsec/pixel), both stars appear blended in our frames. \label{fig13}}
\end{figure}

 \clearpage
\begin{figure}

\epsscale{1.0}
\plotone{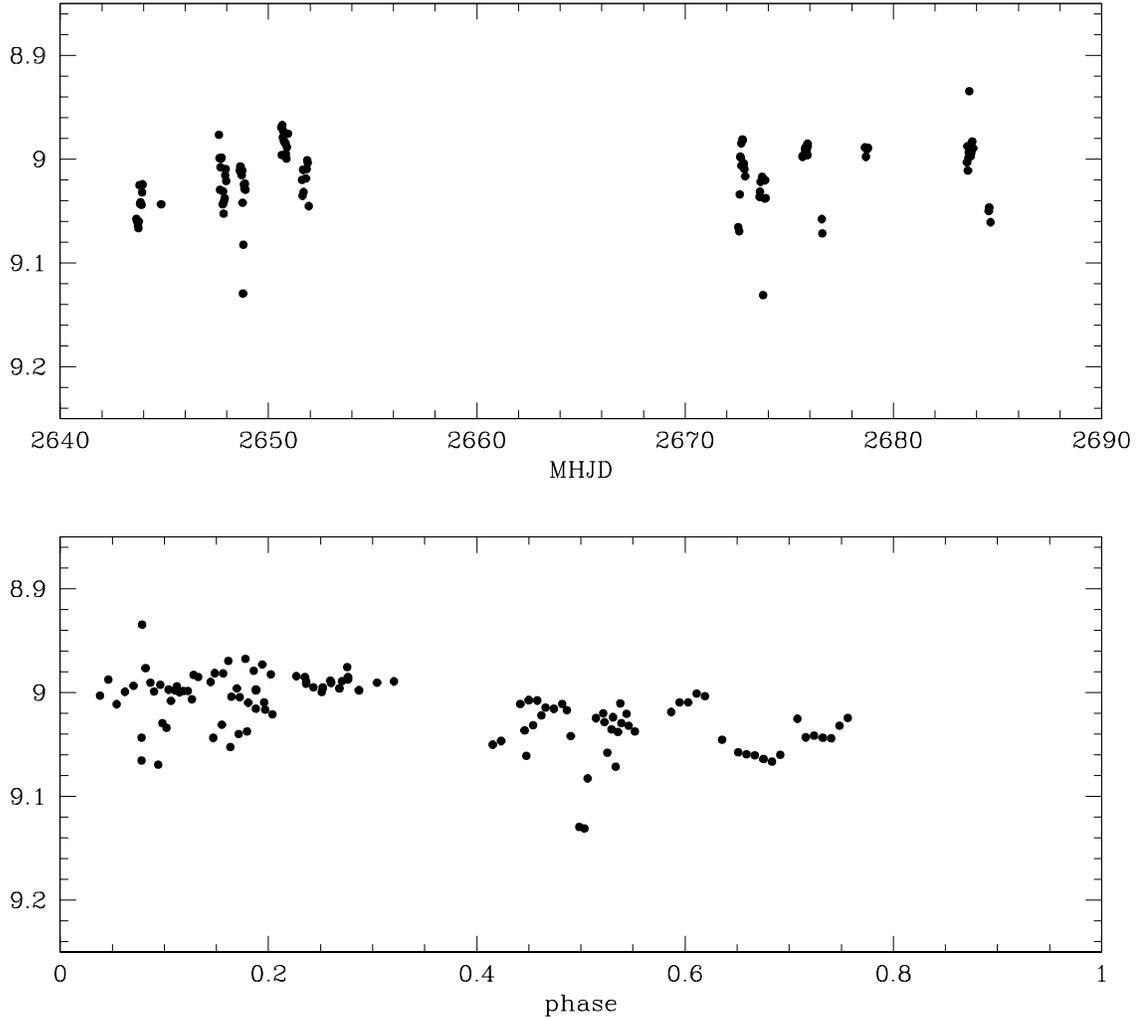}
\caption{(Top) Light curve in Modified Heliocentric Julian Days (+2450000) of the blend between GJ2069A and the brighter nearby star.(Bottom) The same light curve after folding it with the orbital period of 2.771468 days derived by \citet{del99}. See discussion in text.\label{fig14}}
\end{figure}

 \clearpage
\begin{figure}
\epsscale{1.0}
\plotone{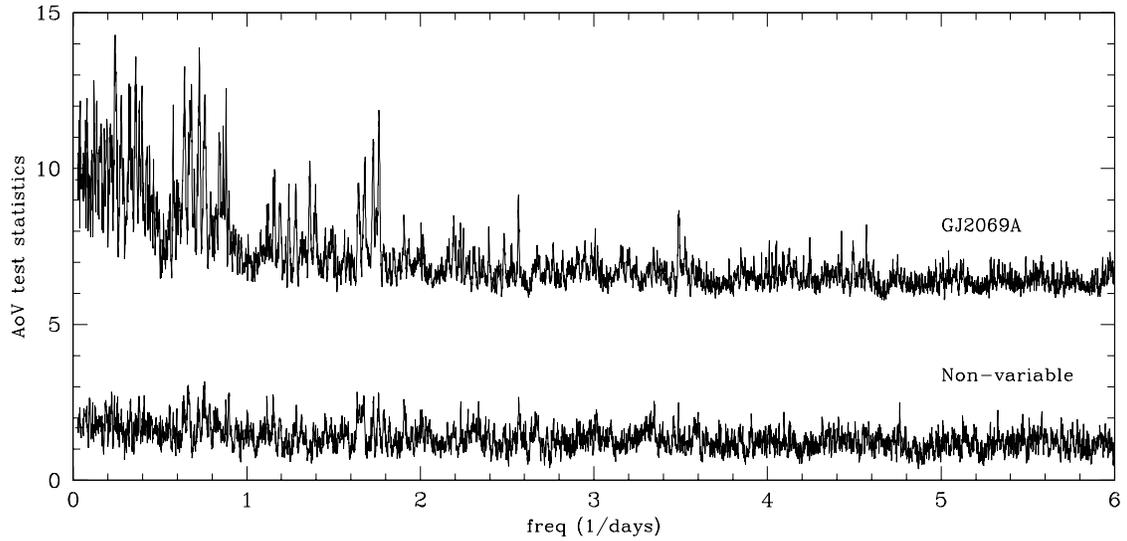}
\caption{ AoV Periodograms derived for the M-dwarf binary GJ 2069A (top), and a non-variable star (bottom). The AoV periodogram of the non-variable is featureless, while several variance peaks are visible in the one for the binary. The peak at $\sim$ 0.36 $days^{-1}$ reproduces the orbital period of the binary derived by \citet{del99}. \label{fig15}}
\end{figure}






\clearpage

\begin{table}
\caption{Summary of the location of the Pisgah Survey and its main hardware\label{tbl-1}}
\footnotesize
\vspace{3mm}
\begin{tabular}{|l|cccc|}	
\tableline\tableline
{\it Location:}&&&&\\
PARI\tablenotemark{a}&{\it Latitude}:&{\it Longitude}:&{\it Elevation}:&\\ 
&$35^{\circ}$14'N&$82^{\circ}$44'W &$\sim$ 600 m&\\ \hline
{\it Telescope:}&&&&\\
Meade Schmidt-Cass.&{\it Diameter:}& {\it Focal ratio:}& {\it Plate scale:}& \\
& 0.2 m & f/6.3  & 161.14 ``/mm &\\ \hline 
{\it Mount:}&&&&\\
Paramount G1100-S&German-&{\it Slew rate:}&&{\it Control Software:} \\
from Software Bisque&Equatorial&5.0 deg$\cdot$$sec^{-1}$&&TheSky\tablenotemark{b}\\ \hline 
{\it Camera:}&&&&\\
Ap-10 Apogee CCD& 2048 x 2048&{\it FOV:}&{\it Pixel scale:}&{\it Control Software:}\\ 
&14$\mu$m pixels&1.647 sq. deg.&2.256 ``/pix&CCDSoft\tablenotemark{b}\\ \hline
{\it Filter:}&&&&\\
50mm-round Bessel filter& ${\lambda}_{center}$:& Bandwidth (${\Delta}$${\lambda}_{FWHM}$):& {\it Peak efficiency:}&\\ 
from Omega Optical Inc.&8000 $\AA$& $\sim$ 1500 $\AA$& $\sim$ 97 $\%$&\\ \hline
{\it Dome:}&&&&\\
HD-10 dome from &{\it Diameter:}&{\it Shutter width:}&{\it Slew rate:}&{\it Control Software:}\\
Technical Innovations Inc.& 3.05 m&0.91 m&$\sim$ 6 deg/sec&Digital Domeworks\tablenotemark{c}\\ 
\tableline\tableline
\end{tabular}
\tablenotetext{a}{Pisgah Astronomical Research Institute, Rosman, NC. http://www.pari.edu}
\tablenotetext{b}{Professional Astronomy Software Suite, Software Bisque Inc.}
\tablenotetext{c}{Technical Innovations Inc.}
\end{table}

\clearpage
\begin{table}
\begin{center}
\caption{Estimated cost of the commercial items used in the construction of the Pisgah Survey \label{tbl-2}}
\vspace{3mm}
\begin{tabular}{|cr|}	
\tableline\tableline
 Part & Estimated cost\tablenotemark{*}\\
\tableline
Telescope&3,500\\
Mount&10,000\\
CCD&14,000\\
Filter&200\\
Filter-camera adapter&225\\
Dome and enclosure\tablenotemark{**}&48,000\\
Computers&3,500\\
Software&1,800\\
\tableline
TOTAL:& 81,225\\
\tableline\tableline
\end{tabular}
\tablenotetext{*}{The prices are in dollars and correspond to purchases made between 1999 and 2002}
\tablenotetext{**}{This includes parts cost, site preparation, cabling, construction, maintenance labor, etc ...}
\end{center}
\end{table}

\clearpage
\begin{deluxetable}{cc}
\tabletypesize{\footnotesize}
\tablewidth{0pt}
\tablecaption{Average I-band sky brightness at PARI from August to November 2000. 
The uncertainties reflect the variation of the sky brightness during each month \label{tbl-3}}

\tablehead{
\colhead{Month (2000)}&\colhead{Sky Brightness (mag/sq. arcsec)}
}
\startdata
August&15.3 $\pm$ 0.7\\
September&15.6$\pm$ 0.6\\
October&16.2$\pm$ 0.5\\
November&16.2$\pm$ 0.5\\
\enddata
\end{deluxetable}

\clearpage
\begin{table}
\begin{center}
\caption{Central coordinates, number of measurements, number of stars per field, and number of bona fide and suspected variables detected in the first 9 fields monitored by the Pisgah Survey. \label{tbl-4}}
\vspace{3mm}
\begin{tabular}{cccccc}
\tableline\tableline
Field ID & $RA_{center}$(J2000) & $Dec_{center}$(J2000) & $N_{measures}$ & \tablenotemark{a} $N_{stars}$ & $N_{vars}$ \\
\tableline
PS-1&$08^{h}31^{m}$ & $19^{\circ}24\arcmin$ & 118 & 843 & 5\\
PS-2&$08^{h}31^{m}$ & $18^{\circ}12\arcmin$ & 106 & 890 & 0\\
PS-3&$08^{h}31^{m}$ & $17^{\circ}00\arcmin$ & 129 & 953 & 4\\
PS-4&$08^{h}31^{m}$ & $15^{\circ}48\arcmin$ &  82 & 805 & 1\\
PS-5&$08^{h}31^{m}$ & $14^{\circ}36\arcmin$ & 126 & 946 & 0\\
PS-6&$08^{h}26^{m}$ & $19^{\circ}24\arcmin$ & 124 & 927 & 2\\
PS-7&$08^{h}26^{m}$ & $18^{\circ}12\arcmin$ & 119 & 903 & 2\\
PS-8&$08^{h}26^{m}$ & $17^{\circ}00\arcmin$ & 132 & 963 & 3\\
PS-10&$08^{h}26^{m}$ & $14^{\circ}36\arcmin$ & 130 & 971 & 3\\
\tableline
Total: & & & &8201 & 20\\
\tableline
\tablenotetext{a}{Only stars with more than 50 detections}
\end{tabular}
\end{center}
\end{table}

\clearpage
\begin{deluxetable}{ccccccccccc}
\tabletypesize{\scriptsize}
\rotate
\tablewidth{0pt}
\tablecaption{Summary of the main parameters of the 20 variable candidates identified in the first 14.76 sq. degrees of sky monitored by the Pisgah Survey.  \label{tbl-4}}
\tablehead{
\colhead{ID}&\colhead{RA(J2000)}&\colhead{$\sigma_{RA}$(sec)}&\colhead{Dec(J2000)}&\colhead{$\sigma_{Dec}$(arcsec)}&\colhead{$\overline{I}$}&\colhead{$\Delta$I(mags)}&\colhead{Period(days)}&\colhead{$T_{o}$(+2450000)}&\colhead{Type\tablenotemark{*}}&\colhead{cross-ID}}
\startdata
PS-1-vs0013&08:32:02.40&$\pm 0.03$&+19:36:01.77&$\pm 0.25$&9.87&$\sim$ 0.1&&&Unknown&--\\
PS-1-vs0020&08:31:37.62&$\pm 0.04$&+19:23:40.82&$\pm 0.30$&9.01&$\sim$ 0.20&&&Unknown&GJ 2069A\\
PS-1-vs0038&08:32:25.48&$\pm 0.10$&+19:56:12.95&$\pm 1.42$&11.99&$\sim$ 0.25&1.51976 $\pm 0.00231$&2684.66845&SR&--\\
PS-1-vs0084&08:30:52.86&$\pm 0.02$&+19:17:37.36&$\pm 0.24$&11.06&$\sim$ 0.1&&&Unknown&NSV 4110\\
PS-1-vs0218&08:28:22.59&$\pm 0.04$&+19:10:07.24&$\pm 0.45$&12.64&$\sim$ 0.55&0.50891 $\pm 0.00026$&2650.88891&EW&--\\
PS-3-vs0011&08:29:39.30&$\pm 0.11$&+17:17:01.10&$\pm 0.62$&10.28&$\sim$ 0.57&1.32275 $\pm 0.00175$&2643.82173&EA&FF Cnc\\
PS-3-vs0048&08:31:23.08&$\pm 0.05$&+17:28:47.84&$\pm 2.98$&10.62&$\sim$ 0.42&&&Long-Term&--\\
PS-3-vs0111&08:31:10.08&$\pm 0.03$&+16:39:18.95&$\pm 0.43$&11.42&$\sim$ 0.18&&&Unknown&--\\
PS-3-vs0116&08:30:44.84&$\pm 0.03$&+16:34:41.95&$\pm 0.42$&10.98&$\sim$ 0.12&3.04878 $\pm 0.06648$&2684.60578&EA&--\\
PS-4-vs0386&08:31:10.08&$\pm 0.03$&+16:39:18.95&$\pm 0.43$&11.42&$\sim$ 0.18&&&Unknown&--\\
PS-6-vs0026&08:27:40.57&$\pm 0.01$&+19:15:44.39&$\pm 0.24$&9.49&$\sim$ 0.25&&&Long-Term&GV Cnc\\
PS-6-vs0131&08:26:35.36&$\pm 0.02$&+18:51:57.57&$\pm 0.29$&11.67&$\sim$ 0.14&0.47961 $\pm 0.00023$&2673.59788&RR/EB&NSV 4077\\
PS-6-vs0262&08:28:22.50&$\pm 0.05$&+19:10:08.22&$\pm 0.61$&12.61&$\sim$0.42&0.16963 $\pm 0.00003$&2647.73606&EW&--\\
PS-7-vs0277&08:27:13.57&$\pm 0.03$&+17:40:35.76&$\pm 0.69$&12.28&$\sim$ 0.40&1.62866 $\pm 0.00265$&2650.83929&E&--\\
PS-7-vs0414&08:24:05.50&$\pm 0.07$&+17:44:04.79&$\pm 1.14$&13.37&$\sim$ 0.28&0.93985 $\pm 0.00088$&2648.73039&E&--\\
PS-8-vs0031&08:23:58.30&$\pm 0.02$&+17:06:50.91&$\pm 0.71$&9.70&$\sim$ 0.39&1.79856 $\pm 0.00649$&2651.83420&EA&--\\
PS-8-vs0062&08:26:23.44&$\pm 0.03$&+17:23:22.32&$\pm 0.60$&11.76&$\sim$ 0.20&21.27660 $\pm 0.52020$&--&CEP&--\\
PS-8-vs0323&08:28:16.67&$\pm 0.07$&+17:26:53.43&$\pm 0.93$&13.35&$\sim$ 0.40&&&Unknown&--\\
PS-10-vs0027&08:28:05.36&$\pm 0.02$&+14:34:55.56&$\pm 0.30$&9.99&$\sim$ 0.09&&&Unknown&--\\
PS-10-vs0189&08:26:45.87&$\pm 0.03$&+14:58:21.90&$\pm 0.64$&12.64&$\sim$ 0.25&0.94073 $\pm 0.00088$&2650.85500&E&--\\\\
\enddata
\tablenotetext{*}{CEP=Cepheid; E=Eclipsing binary system; EA= Algol-type eclipsing system; EB= Beta Lyrae-type eclipsing system; EW= W Ursae Majoris-type eclipsing variable; RR= RR Lyrae variable; SR= Semiregular variable}
\end{deluxetable}

\begin{deluxetable}{ccccc}
\tabletypesize{\footnotesize}
\tablewidth{0pt}
\tablecaption{Cataloged variables not identified by our variability search. \label{tbl-5}}

\tablehead{
\colhead{RA(J2000)}&\colhead{Dec(J2000)}&\colhead{Type}&\colhead{mag(V)}&\colhead{Field ID}
}
\startdata
08:31:39.10&+18:06:13&Unknown&11.50&PS-2\\
08:30:50.20&+14:22:26&Unknown&9.8&PS-5\\
08:26:39.50&+19:03:28&S\tablenotemark{*}&12.40&PS-10\\
\enddata
\tablenotetext{*}{unstudied variable star with rapid light changes \citep{kho98}}
\end{deluxetable}

\end{document}